\documentclass{elsart}

\usepackage{epsfig}

\usepackage{amssymb}

\begin{document}
\begin{frontmatter}

\title{The additivity of the pseudo-additive conditional entropy\\
for a proper Tsallis' entropic index}
\author[wada]{Wada Tatsuaki\corauthref{cor1}} and
\ead{wada@ee.ibaraki.ac.jp}
\author[saito]{Saito Takeshi}
\ead{saito@kif.co.jp}
\address[wada]{Department of Electrical and Electronic Engineering, 
Ibaraki University, Hitachi,~Ibaraki, 316-8511, Japan}
\corauth[cor1]{Corresponding author.}
\address[saito]{KIF \& Co., Ltd., Tokyo Dia Bldg. \#5, 
15 floor, 1-28-23 Shinkawa Chuo-ku, Tokyo, 104-0033, Japan}

\begin{abstract}
For Tsallis' entropic analysis to the time evolutions of standard logistic map 
at the Feigenbaum critical point, 
it is known that there exists
a unique value $q^*$ of the entropic index such that the asymptotic
rate $K_q \equiv \lim_{t \to \infty} \{S_q(t)-S_q(0)\} / t$ of increase 
in $S_q(t)$ remains finite whereas $K_q$ vanishes (diverges) for 
$q > q^* \; (q < q^*)$.
We show that in spite of the associated
whole time evolution cannot be factorized into a product of independent 
sub-interval time evolutions, the pseudo-additive
conditional entropy 
$S_q(t|0) \equiv \{S_q(t)-S_q(0)\}/ \{1+(1-q)S_q(0)\}$ becomes 
additive when $q=q^*$. 
The connection between $K_{q^*}$ and the rate $K'_{q^*} \equiv 
S_{q^*}(t \vert 0) / t$ of increase in the conditional entropy is discussed.
\end{abstract}

\end{frontmatter}

\section{Introduction}
There has been growing interest on the non-extensive formalism based on
Tsallis' entropy \cite{Tsal88,Tsal01}
\begin{equation}
S_q \equiv \frac{1 - \sum_i p_i^q}{q - 1},
\end{equation}
\noindent
which is characterized by the index $q$. In the limit $q \to 1$, 
$S_q$ reduces to Boltzmann entropy.
In order to deeply understand what the non-extensive formalism means, 
the mixing properties of one dimensional (1D) dissipative maps 
\cite{Tsal01,Cost97,Lyra98,Lato00,Mour00} have been studied based on $S_q$.
It is shown that there exists 
the special value $q^*$ of entropic index of $S_q(t)$ such that 
$K_q \equiv \lim_{t \to \infty} \{S_q(t)-S_q(0)\} / t$
is finite for $q=q^*$, and vanishes (diverges) for $q>q^* (q<q^*)$.
We here focus on the existence of such a unique value of $q^*$ 
since it may shed some light on the meaning of $q$
in the non-extensive formalism.  
Latora {\it et al.} \cite{Lato00} have found that $S_q(t)$ linearly grows 
when the entropic index $q$ equals $q^* \simeq 0.2445$ for 
standard logistic map at the Feigenbaum critical point.
It is also known \cite{Tsal01,Cost97,Lyra98,Lato00} that the  
$q^*$ can be obtained from three different methods: 
i) the upper-bound of a sensitivity of initial conditions
 $\quad \xi(t) \propto t^{\frac{1}{1-q^*}}$;
ii) the singularity strength of a multi-fractal distribution
$(1-q^*)^{-1} = \alpha_{\rm min}^{-1} - \alpha_{\rm max}^{-1}$;
iii) Tsallis' entropy becomes linear in time
$\quad S_{q^*} (t) \propto t$.
In addition all three methods yield the same value of $q^* \simeq 0.2445$ 
for standard logistic map.

We have proposed an alternative method \cite{Wada01} of obtaining $q^*$ 
as follows:
the conditional non-extensive entropy which is introduced by Abe and Rajagopal
\cite{Abe00},
\begin{equation}
S_q(t \vert 0) \equiv \frac{S_q(t) - S_q(0)}{1 + (1-q) S_q(0)},
\label{cond-Ent}
\end{equation}
\noindent
becomes {\it additive} if and only if $q$ equals $q^*$.
We here explain the underlying simple mechanism in our method 
of obtaining $q^*$. Then the works of Latora {\it et al.} \cite{Lato00} and 
de Moura {\it et al.} \cite{Mour00} are explained in terms of our method. 
The connection between the rates $K_{q^*}$ and $K'_{q^*} \equiv 
S_{q^*}(t \vert 0) / t$ is also discussed.

\section{The conditional Tsallis' entropy method of obtaining $q^*$}
Let us begin with explaining the essence of our method \cite{Wada01} of 
obtaining $q^*$. 
Suppose the number of states $W(t)$ obeys a power-law evolution
expressed with $W(t) = W(t_0) \exp_{q^*}( K'_{q^*} \; (t-t_0) )$ and 
$S_q(t) = \ln_q W(t)$, where $\exp_q(x) \equiv [1+(1-q)x]^{\frac{1}{1-q}}$ and 
$\ln_q x \equiv (x^{1-q}-1)/(1-q)$.
Then the corresponding conditional Tsallis' entropy Eq. (\ref{cond-Ent}) can 
be written as
\begin{equation}
  S_q(t|0) =  \ln_q [\frac{W(t)}{W(0)}] 
           = \ln_q[\; \exp_{q^*}(K'_{q^*} \; t) \;],
\end{equation}
\noindent
where the relation $\ln_q[y/x] = (\ln_q y - \ln_q x)/x^{1-q}$ is used.
From the last expression, we see that $S_q(t \vert 0)$ is proportional to $t$
if and only if $q = q^*$, consequently $S_{q^*}(t \vert 0)$ becomes additive, 
i.e., $S_{q^*}(t_1 + t_2 \vert 0) = S_{q^*}(t_1 \vert 0) +
                                  S_{q^*}(t_2 \vert 0)$.

We now explain the work of Latora {\it et al.} \cite{Lato00} in the view of 
our method. In their work the whole phase space ($-1 \le x \le 1$)
was divided into a large number $w$ of equal cells and 
the time evolution of $S_q(t)$ for standard logistic map was calculated. 
The key points of their procedure are:
i) their initial distribution is that all $N$ points are in a single 
   cell $x(0)$ at $t=0$, consequently $S_q(0)=0$; and
ii) averaging $S_q(t)$ over the best initial cells $\{x(0)\}$.
Note that since $S_q(0) = 0$, the conditional Tsallis' entropy $S_q(t \vert 0)$
reduces to $S_q(t)$. This seems trivial at glance. However the essential point 
is the averaging over the best initial ensemble in ii).  
Our opinion is that what is calculated in their procedures is
nothing but the conditional Tsallis' entropy, which is originally
defined \cite{Abe00} by 
\begin{equation}
  S_q(t|0) \equiv \frac{\sum_{x(0)} p(x(0))^q s_q(t|x(0))}
                         {\sum_{x(0)} p(x(0))^q}.
 \label{cond-Sq} 
\end{equation}
In other words, $S_q(t|0)$ is the $q$-average of the conditional 
entropy $s_q(t|x(0))$ over the initial cells $\{x(0)\}$.
In fact the $p_i(t) = N_i(t) / N $considered in their paper is 
a probability of finding a point in the $i$-th cell $x_i(t)$ 
at time $t \ge 0$ for the initial condition that the only $x(0)$ is 
occupied at $t=0$. 
Hence the $p_i(t)$ is nothing but the conditional probabilities 
$p(x_i(t)|x(0))$! Then introducing $s_q(t|x(0))$ as
\begin{equation}
  s_q(t|x(0)) = \frac{1 - \sum_{i=1}^{w} p(x_i(t)|x_0)^q}{q-1},       
\end{equation}
we can readily show the quantity $S_q(t)$ in their paper
is nothing but $S_q(t|t_0)$ of Eq. (\ref{cond-Sq}). 
In addition, let $M$ be the total number of the best initial cells 
$\{ x(0) \}$ for the averaging, then $p(x(0))=1/M$. 
The Eq. (\ref{cond-Sq}) is rewritten by
\begin{equation}
  S_q(t|0) = \frac{ (1/M)^q \sum_{x(0)}  s_q(t|x(0))}{M (1/M)^q} 
             = \frac{1}{M} \sum_{x(0)=1}^{M}  s_q(t|x(0)),
\end{equation}
which is the conventional average of the conditional Tsallis' entropy 
of $s_q(t|x(0))$ over the best initial cells $\{ x(0) \}$.

We next focus on the work of de Moura {\it et al.} \cite{Mour00}.
They have studied the rate of the convergence to the critical attractor 
of the generalized logistic map. Their initial ensemble consists of $N$ points
that uniformly spread over the entire phase space. 
They assume that $S_{q^*}(t) - S_{q^*}(0)$ evolves at a constant 
rate $K_{q^*}$, and consequently the number
of occupied cells $W(t)$ evolves in time as
\begin{equation}
  W(t) = [W(0)^{1-q^*} + (1-q^*)K_{q^*} t]^{\frac{1}{1-q^*}}.                
\end{equation}
This can be rewritten as
\begin{equation}
  W(t) = W(0) \cdot \exp_{q^*}[ \frac{K_{q^*} t}{ W(0)^{1-q^*}}] 
       = W(0) \cdot \exp_{q^*}[ K'_{q^*} t ],
  \label{W}   
\end{equation}
where $K'_{q^*} = K_{q^*}/W(0)^{(1-q^*)}$.
Note that the relation between $K'_{q^*}$ and $K_{q^*}$ is the same relation 
$ \lambda' = \frac{\lambda}{\bar{Z}_q^{1-q}}$ between the Langrage multiplier 
$\lambda'$ of the optimal Lagrange multiplier (OLM) method \cite {OLM} and 
that $\lambda$ of 
Tsallis-Mendes-Plastino in canonical ensemble formalism.
Since the conditional Tsallis' entropy 
$S_q(t|0)$ is expressed as $S_q(t|0) = \ln_q [W(t)/W(0)]$,
Eq. (\ref{W}) means $S_q(t|0)$ is proportional to $t$ when $q=q^*$.
Hence the both works of Latora et al. \cite{Lato00} and of de Moura et al. 
\cite{Mour00} can be explained in terms of our method of obtaining $q^*$.

\section{Conclusions}
  We have explained that a new kind of additivity of the conditional Tsallis' 
entropy holds for the proper entropic index $q^*$. We also show the works 
of both Latora et al. \cite{Lato00} and of de Moura et al. \cite{Mour00} can 
be explained in terms of our method of obtaining $q^*$. 


\end{document}